\documentclass[12pt]{iopart}

\usepackage{color}
\usepackage{graphicx}
\usepackage{caption}
\usepackage{subcaption}
\usepackage{rotating}
\usepackage{hyperref}
\usepackage{color, soul}

\begin{document}
%\TDM

\title{Graphene Nanoribbon as an Elastic Damper}

\author{Iman Evazzade}
\address
{Department of Physics, Faculty of Science, Ferdowsi University of Mashhad, Mashhad, Iran}
\ead{i.evazzade@mail.um.ac.ir}

\author{Ivan P. Lobzenko}
\address
{Institute for Metals Superplacticity Problems RAS, Ufa 450001, Russia}
\address
{Toyota Technological Institute, 2-12-1 Hisakata, Tempaku-ku, Nagoya, 468-8511 Japan}

\author{Danial Saadatmand}
\address
{Department of Physics, University of Sistan and
Baluchestan, Zahedan, Iran}

\author{Elena A. Korznikova}
\address
{Institute for Metals Superplacticity Problems RAS, Ufa 450001, Russia}

\author{Kun Zhou}
\address
{School of Mechanical and Aerospace Engineering, Nanyang Technological University, 639798, Singapore}
\address
{Environmental Process Modeling Center, Nanyang Environment and Water Research Institute, Nanyang Technological University, 50 Nanyang Avenue, 639798, Singapore}

\author{Bo Liu}
\address
{School of Mechanical and Aerospace Engineering, Nanyang Technological University, 639798, Singapore}
\address
{Environmental Process Modeling Center, Nanyang Environment and Water Research Institute, Nanyang Technological University, 50 Nanyang Avenue, 639798, Singapore}

\author{Sergey V. Dmitriev}
\address
{Institute for Metals Superplacticity Problems RAS, Ufa 450001, Russia}
\address
{National Research Tomsk State University, Tomsk 634050, Russia}

\begin{abstract}
Heterostructures composed of dissimilar two-dimensional nanomaterials can have nontrivial physical and mechanical properties promising for many applications. Interestingly, in some cases, it is possible to create heterostructures composed of weakly and strongly stretched domains with the same chemical composition, as it has been demonstrated for some polymer chains, DNA, and intermetallic nanowires supporting this effect of two-phase stretching. These materials at relatively strong tension forces split into domains with smaller and larger tensile strain. Within this region, average strain increases at constant tensile force due to the growth of the domain with larger strain in expense of the domain with smaller strain. Here the two-phase stretching phenomenon is described for graphene nanoribbons with the help of molecular dynamics simulations. This unprecedented feature of graphene revealed in our study is related to the peculiarities of nucleation and motion of the domain walls separating the domains with different elastic strain. It turns out that the loading-unloading curves exhibit a hysteresis-like behavior due to the energy dissipation during the domain wall nucleation and motion. Here, we originally put forward the idea of implementing graphene nanoribbons as elastic dampers, efficiently converting mechanical strain energy into heat during cyclic loading-unloading through elastic extension where domains with larger and smaller strain coexist. Furthermore, in the regime of two-phase stretching, graphene nanoribbon is a heterostructure for which the fraction of domains with larger and smaller strain, and consequently its physical and mechanical properties, can be tuned in a controllable manner by applying elastic strain and/or heat.
\end{abstract}
\maketitle

\section{Introduction}
Two-dimensional (2D) nanomaterials, due to combination of their unusual mechanical and physical properties, have attracted great attention of researchers in the last decade (see \cite{NewReview} and references therein). Combination of dissimilar 2D materials in various heterostructures is another way to achieve new properties and find new applications. Such heterostructures can be in a number of different ways including through chemical modification, partial hydrogenation of graphene \cite{Melnick2016,Yang2010,Xiang20094025,Liu201418180}, by assembling layered heterostructures using weak van der Waals interactions \cite{Geim2013419,Novoselov2016,Li2017_031005,Zhang2017_035404,Chen2017399}, or by edge-to-edge stacking of dissimilar 2D materials \cite{Liu201418180,Chen2017,Liu2014236,Kistanov2017}. 

\begin{figure}
\begin{center}
\includegraphics[width=1.0\textwidth]{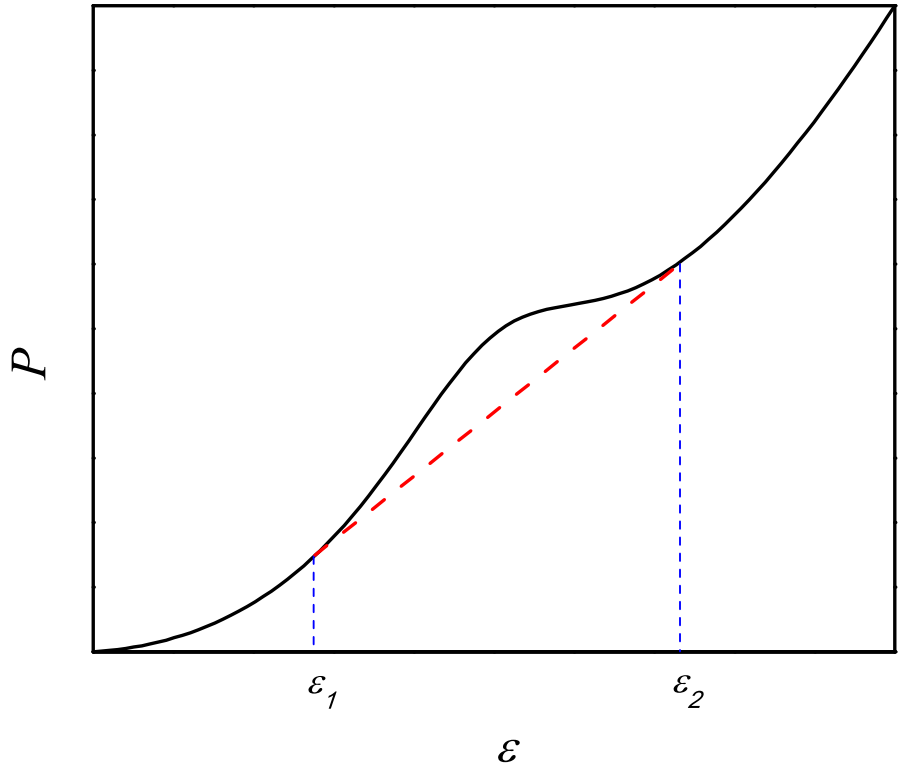}
\caption{Schematic dependence of the potential energy of a translational unit of a material as a function of tensile strain, featuring the concave down region in the range $\varepsilon_1<\varepsilon<\varepsilon_2$ (solid line) \cite{SavinDNA}. Presence of such concave down region on the $P(\varepsilon)$ function is the necessary condition for the two-phase stretching. Within the range $\varepsilon_1<\varepsilon<\varepsilon_2$ there exists a path with a lower potential energy, shown by the straight dashed line, which is tangent to the $P(\varepsilon)$ curve at the points $\varepsilon_1$ and $\varepsilon_2$. When this path is followed, the material splits into domains with tensile strain $\varepsilon_1$ and $\varepsilon_2$, and stretching occurs due to the growth of the domain with larger strain in expense of the domain with smaller strain. Outside the range $\varepsilon_1<\varepsilon<\varepsilon_2$ stretching is homogeneous.}
\label{Fig_1}
\end{center}
\end{figure}

Synthesis of heterostructures is not an easy task \cite{Novoselov2016,Ai20173413,Luo20166904,Zhang2015410,Tongay20143185} and it would be helpful to find new methods of their creation. Savin et al. \cite{SavinDNA} have uncovered a general mechanism of two-phase stretching experimentally observed for DNA \cite{Ikai2016133,King20133859,Zhang20133865,Smith1996795}, polypeptides \cite{Afrin20091105}, and some polymer chains \cite{SavinDNA}. There exists numerical evidence that NiAl, FeAl, and CuZr intermetallic nanowires also split into domains with smaller and larger strain under tension  \cite{Babicheva2013,Bukreeva2013,Bukreeva2014,Bukreeva201391,Sutrakar2010679,Sutrakar20101565}. Depending on loading scheme, temperature and other parameters, reversible phase transitions can be expected in graphene \cite{PT1,PT2,PT3,PT4}, gamma-boron \cite{PT_GB}, and MoS$_2$ \cite{PT_MoS2_1,PT_MoS2_2,PT_MoS2_3}. In the regime of two-phase stretching, a heterostructure naturally appears as a result of application of strain-controlled tension to a material having specific features described in the work \cite{SavinDNA}. In spite of the fact that the domains have same chemical composition, due to considerable difference in their elastic strain, they can differ by lattice symmetry and interatomic distances, resulting in different physical and mechanical properties. This approach to heterostructure creation can be called elastic strain engineering \cite{He2016,Yang20164028,Kalikka2016,Quereda20162931,Si20163207,Xie2016,Qi20121224,Li2014108,Qi20121224,Zhu2010710}, where elastic strain or strain gradient is applied to a material to modify and improve its properties. Nanomaterials can withstand large elastic strain (of the order of 0.01 or even 0.1) before defect formation or fracture begins; that is why, their properties can be considerably altered by the elastic strain \cite{ Li2014108,Baimova2012}.
What is the necessary condition for two-phase stretching? This is explained by considering the dependence of the potential energy of a translational unit of the studied material, $P$, as a function of tensile strain, $\varepsilon$, \cite{SavinDNA}. For many materials this dependence is concave upward, but for some of them it may feature a concave down region in the range of tensile strain $\varepsilon_1<\varepsilon<\varepsilon_2$, as schematically shown in Fig.~\ref{Fig_1} by the solid line. Presence of such concave down region on the $P(\varepsilon)$ function is the necessary condition for the two-phase stretching. Within the range of tensile strain $\varepsilon_1<\varepsilon<\varepsilon_2$ there exists a path with lower potential energy, shown by the straight dashed line, which is the tangent of the $P(\varepsilon)$ function at the points $\varepsilon_1$ and $\varepsilon_2$. When the system moves along this path, the material splits into domains with tensile strain $\varepsilon_1$ and $\varepsilon_2$, and stretching occurs due to the growth of the domain with larger strain in expense of the domain with smaller strain. Outside the range $\varepsilon_1<\varepsilon<\varepsilon_2$ stretching is homogeneous. Tensile force is constant within the domain of $\varepsilon_1<\varepsilon<\varepsilon_2$, since it is proportional to the derivative of potential energy, $dP/d\varepsilon$, and $P(\varepsilon)$ is linear. 

From the picture described above it can be seen that two-phase stretching occurs through the motion of domain walls (DWs) separating domains with different tensile strain. To have the possibility of witnessing two-phase stretching, the size of the material should be sufficiently large to host at least one DW generated at the surface or at least two DWs (positive and negative) if they are generated in the bulk. Actually, generation and motion of DWs in the two-phase stretching may have interesting consequences on the behavior of materials; e.g. resulting in negative stiffness within the strain range $\varepsilon_1$ and $\varepsilon_2$ \cite{Bukreeva2013,Bukreeva2014} or in strong hysteresis of the stress-strain response \cite{PT1,PT3,PT_MoS2_2}.

In the present study, with the help of molecular dynamics simulations, we demonstrate that the armchair graphene nanoribbons go through a two-phase stretching. Furthermore, we present that the nucleation and motion of DWs produces considerable dissipation of elastic strain energy so that the nanoribbon combines the properties of elasticity and damping. A similar combination of elasticity and damping demonstrates such macroscopic materials as rubber or porous metallic rubber, finding many engineering applications \cite{Paimushin2016435,Ponomarev20138}. To the best of our knowledge, such analogs have not been proposed for 2D nanomaterials.

\begin{figure}
\begin{center}
\includegraphics[width=1.0\textwidth]{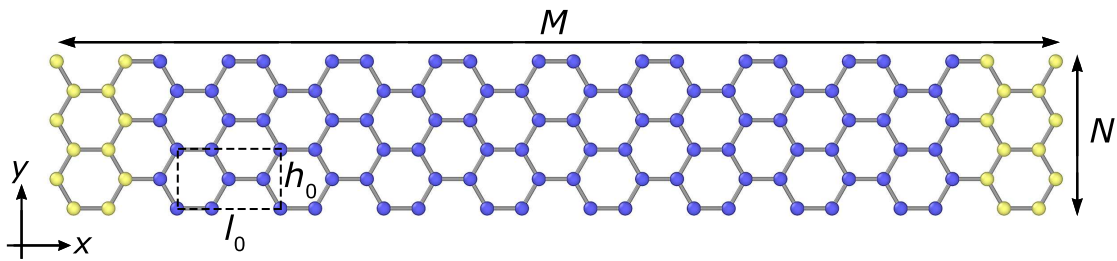}
\caption{Graphene nanoribbon structure. The $x$ ($y$) axis is along the armchair (zigzag) direction. The rectangular translational unit cell includes four carbon atoms (shown by the dashed line). The initial nanoribbon length (width) is $L_0=Ml_0$ ($H_0=Nh_0$), where $M$ and $N$ are the numbers of translational cells along the $x$ and $y$ axes, respectively. We take $M=40$ and $N=3$, 6, or 12. The atoms at the nanoribbon ends are clamped (shown in yellow). The edges parallel to the $x$ axis are free. The nanoribbon is subjected to quasi-static stretching (unstretching) along the $x$ axis by a stepwise increase (decrease) of the axial strain followed by structure relaxation after each increment. }
\label{Fig_2}
\end{center}
\end{figure}

\section{Materials and methods}
Uniaxial tension of graphene nanoribbons is studied by molecular dynamic simulations using the LAMMPS program package \cite{lammps} with the use of AIREBO potentials \cite{airebo}. The initial graphene nanoribbon structure is schematically shown in Fig.~\ref{Fig_2}. The armchair (zigzag) graphene direction is along the $x$ ($y$) axes. We consider a rectangular translational cell with four atoms as shown by the dashed line. The initial size of this cell is $l_0$=4.26~\AA and $h_0$=2.46~\AA. To simulate the nanoribbons, a rectangular computational cell consisting of $M\times N$ translational cells is built. The translational cells are thus numbered by the indices $(m,n)$. In Fig.~\ref{Fig_2}, the nanoribbon with $M=10$ and $N=3$ is shown, but in our simulations the computational cell has $M=40$ cells along the $x$ axis and $N=3$, $N=6$, or $N=12$ cells along the $y$ axis. The actual dimensions of the studied nanoribbons are: the three different initial widths $H_0=Nh_0-h_0/2$= 6.07, 13.41, and 28.03 \AA, with the same initial length of $L_0=Ml_0$=166.05 \AA. The total number of atoms in the computational cell is equal to $4MN$. To apply strain-controlled tension, two zigzag chains at each end of the nanoribbon are assumed to be clamped (shown in yellow). The nanoribbon edges parallel to the $x$ axis are assumed to be free. The nanoribbon is subjected to quasi-static stretching (unstretching) along the $x$ axis by a stepwise increase (decrease) of the averaged axial strain followed by structure relaxation after each increment. Strain increment is equal to $\Delta\varepsilon_{xx}=0.01$.  The averaged longitudinal and transverse strains are defined as $\varepsilon_{xx}=(L-L_0)/L_0$ and $\varepsilon_{yy}=(H-H_0)/H_0$, respectively, where $L$ is the current distance between the nanoribbon ends and $H$ is the current nanoribbon width. The local axial strain is calculated for the translational cells with the numbers $m=1,...,M$, $n=N/2$ ($n=2$ for $N=3$) as $e_{xx}(m)=(l_m-l_0)/l_0$, where $l_m$ is the current length of the cell. Most of the results are obtained for zero temperature. However, the effect of temperature is discussed in Sec. Temperature Effect. 

\section{Numerical results}

Here we present the simulation results on stretching/unstretching of graphene nanoribbons schematically shown in Fig.~\ref{Fig_2}.  

\subsection{Relaxational dynamics}

% We employ molecular dynamics to simulate tension of graphene nanoribbons schematically shown in Fig.~\ref{Fig_2}, where the $x$ ($y$) axis is along the armchair (zigzag) direction. Strain-controlled tension along the $x$ axis is achieved by stepwise increasing the distance between nanoribbon edges with clamped atoms (shown in yellow). The nanoribbon edges parallel to the $x$ axis are free. Rectangular translational cell containing four carbon atoms is shown by dashed line. It has dimensions of $l_0$ = 4.26~\AA, $h_0$ = 2.46~\AA. We always take $M=40$ translational cells along the $x$ axis (initial nanoribbon length equals to $L_0 = Ml_0$ = 166.05 \AA) and consider the effect of the nanoribbon width $H_0=Nh_0-h_0/2$ by comparing the results for $N=3$, $N=6$, and $N=12$ translational cells along the $y$ axis. We control average axial strain $\varepsilon_{xx}=(L-L_0)/L_0$, where $L$ is the current nanoribbon length. We also calculate local axial strain for each $m$-th translational cell ($m=1,...,M$) with $n=N/2$ as follows, $e_{xx}(m)=(l_m-l_0)/l_0$, where $l_m$ is the current length of the cell and $l_0=L_0/M$ is its initial length. After each small increment in the average axial strain, $\varepsilon_{xx}$, the structure is subjected to relaxation in order to simulate tension at  0~K (the effect of temperature is also addressed later). More details on the simulation method are given in Sec.~Materials and Methods. 

\begin{figure}
\begin{center}
\includegraphics[width=0.8\linewidth]{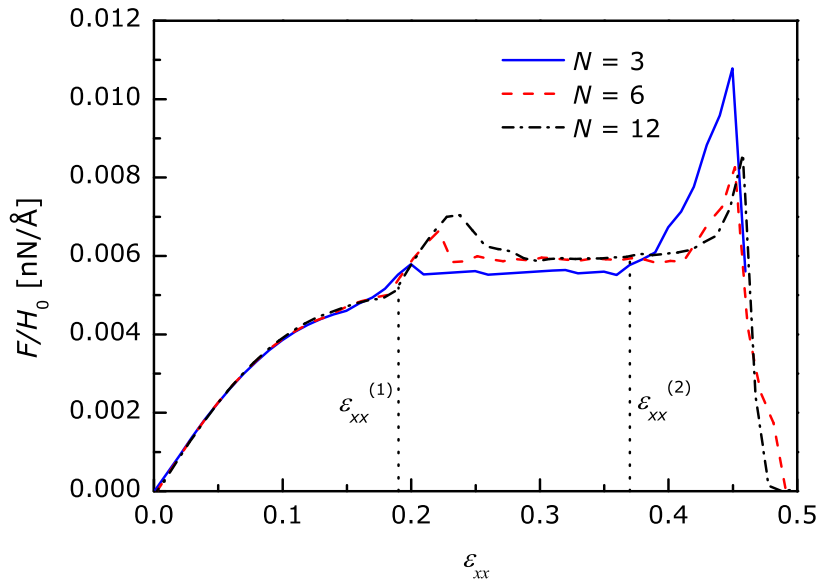}
\caption{Tensile membrane force as a function of averaged tensile strain for the nanoribbons of different widths: $N=3$ (blue solid line), $N=6$ (red dashed line), $N=12$ (black dash-dotted line).}
\label{Fig_3}
\end{center}
\end{figure}

In Fig.~\ref{Fig_3} tensile membrane force as a function of averaged tensile strain is presented for the nanoribbons of different widths: $N=3$ (blue solid line), $N=6$ (red dashed line), and $N=12$ (black dash-dotted line). Note that axial force $F$ applied to the nanoribbon is measured in nanonewtons and normalized by the initial nanoribbon width $H_0$ measured in angstroms. At small strain $(\varepsilon_{xx}<0.05)$, membrane force increases linearly with strain, and then in the domain $0.05<\varepsilon_{xx}<\varepsilon_{xx}^{(1)}$ the tensile stiffness of the nanoribbon gradually reduces with increasing strain. For all the three curves $\varepsilon_{xx}^{(1)}$ equals 0.19 . Then each curve features a hump, whose height increases with increasing nanoribbon width, followed by a plateau at the level of 0.006~nN/\AA. For the narrowest nanoribbon the plateau is at a slightly smaller value due to the more pronounced effect of free edges. The plateau ends at $\varepsilon_{xx}=\varepsilon_{xx}^{(2)}=0.375$ (this value is given for $N=3$ and slightly increases with increasing $N$), and for larger average strain the membrane force increases until it reaches the maximal value (at about $\varepsilon_{xx}$=0.45) where nanoribbon rupture begins resulting in a sudden drop of the membrane force.

\begin{figure}
\begin{center}
\includegraphics[width=0.8\linewidth]{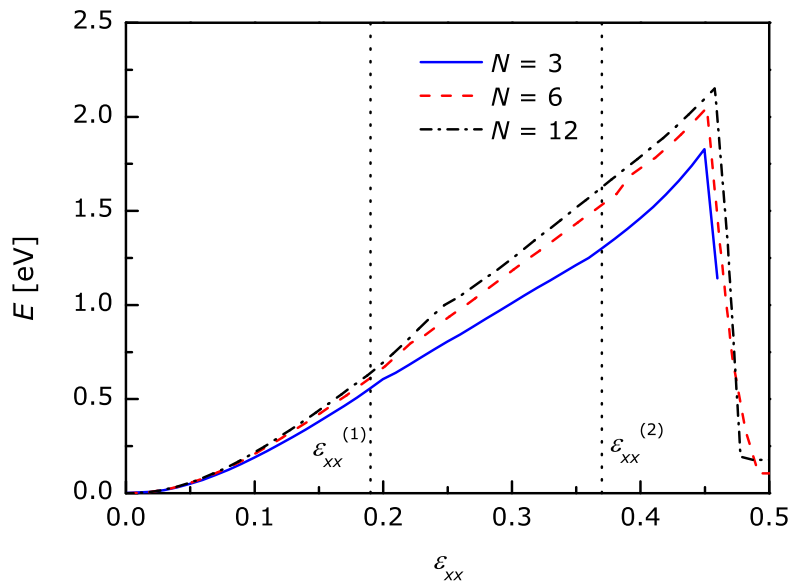}
\caption{ Potential energy per atom as a function of averaged tensile strain for the nanoribbons of different widths:  $N=3$ (blue solid line), $N=6$ (red dashed line), $N=12$ (black dash-dotted line). Sudden drop of energy corresponds to nanoribbon rupture.}
\label{Fig_4}
\end{center}
\end{figure}

\begin{figure}
\begin{center}
\includegraphics[width=0.7\linewidth]{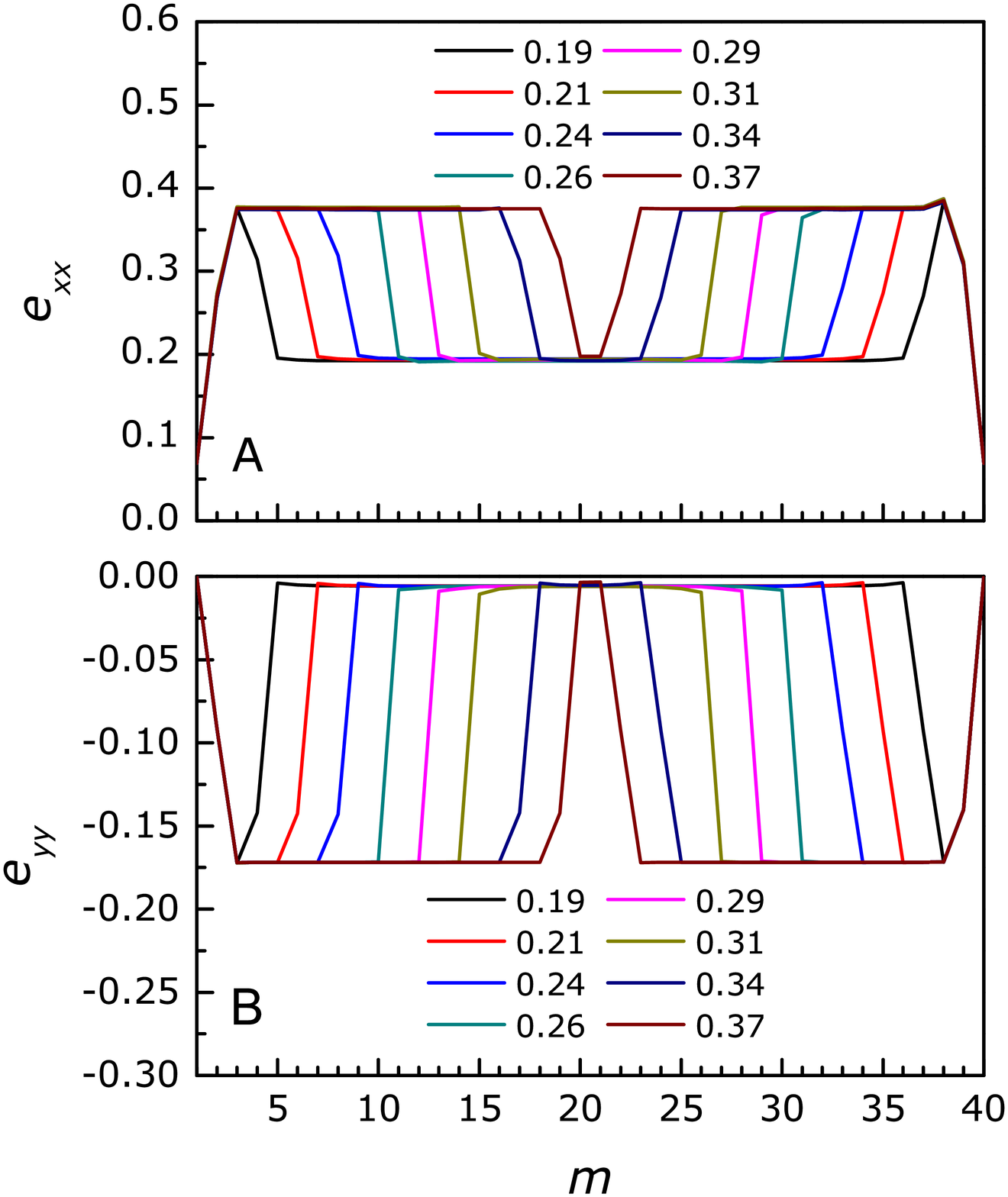}
\caption{Distribution of ($A$) axial and ($B$) lateral strain in the nanoribbon with $N=3$ at different levels of the averaged tensile strain $\varepsilon_{xx}$, indicated in the insets.}
\label{Fig_5}
\end{center}
\end{figure}

Potential energy per atom for the three studied nanoribbons is shown in Fig.~\ref{Fig_4} as a function of the averaged tensile strain. For small strain, one has $E \sim \varepsilon_{xx}^2$ as it should be for the linear elastic body. In the range of averaged tensile strain $\varepsilon_{xx}^{(1)}<\varepsilon_{xx}<\varepsilon_{xx}^{(2)}$ the potential energy growth with strain is nearly liner. At higher strain, it increases faster than linearly until the nanoribbon ruptures with a sudden drop of the energy.
In order to explain the peculiarities of the curves shown in Fig.~\ref{Fig_3} and Fig.~\ref{Fig_4}, the nanoribbon structure is analyzed at different levels of the averaged tensile strain. It is found that in the range of $\varepsilon_{xx}^{(1)}<\varepsilon_{xx}<\varepsilon_{xx}^{(2)}$ the nanoribbon stretching is inhomogeneous. The strain state in this range of averaged strain can be characterized by the components of local strain, $e_{xx}(m)$ and $e_{yy}(m)$, calculated for $m=1,...,M$ and $n=N/2$ (for $N=3$ we choose $n=2$). Results are presented in Fig.~\ref{Fig_5}, where the values of the averaged tensile strain $\varepsilon_{xx}$ are given in the insets for each curve. As one can see from Fig.~\ref{Fig_5}$A$, at $\varepsilon_{xx}=0.19$ the two domains with $e_{xx}=0.375$ appear at the two ends of the nanoribbon and they grow with increasing averaged strain in expense of the domain with $e_{xx}=0.19$. Appearance of the domains at the nanoribbon ends is explained by the clamped boundary conditions producing strain gradients near the ends. For $\varepsilon_{xx}>0.375$ the domain with smaller value of $e_{xx}$ disappears and further stretching is homogeneous until the rupture point. The local lateral strain shown in Fig.~\ref{Fig_5}$B$ is also inhomogeneous within the range of the average strain $\varepsilon_{xx}^{(1)}<\varepsilon_{xx}<\varepsilon_{xx}^{(2)}$. At $\varepsilon_{xx}=0.19$ in the domain of $e_{yy}=-0.01$ the two domains with $e_{yy}=-0.175$ appear near the ends of the nanoribbon and grow with increasing average strain until $\varepsilon_{xx}=0.375$. At higher average strain, stretching is homogeneous. The result is given in Fig.~\ref{Fig_5} for the narrowest nanoribbon ($N=3$) but the same was observed for the two wider nanoribbons.

\begin{figure}
\begin{center}
\includegraphics[scale=0.5]{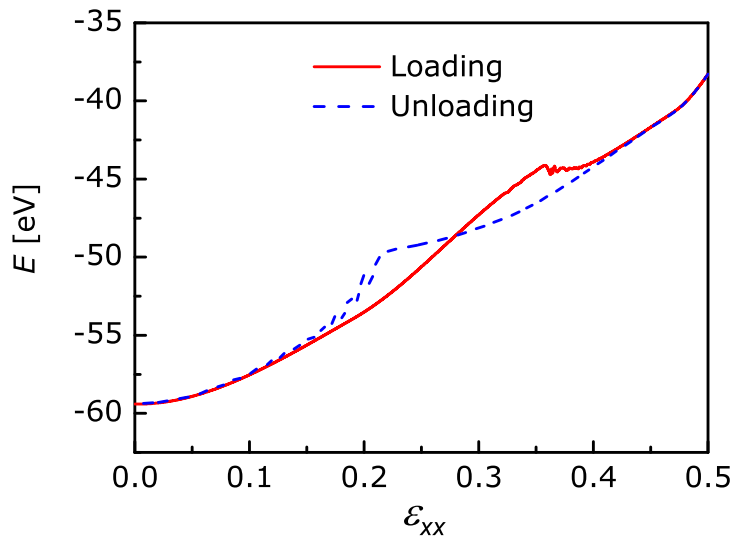}
\caption{Energy as the function of strain $\varepsilon_{xx}$ for the translational cell containing four atoms, as shown in Fig.~\ref{Fig_2}. Results for loading and unloading are shown by red solid and blue dashed curves, respectively. The concave down regions, responsible for the two-phase stretching of nanoribbons, are clearly seen for both curves. The lower energy path along the tangent straight line, as schematically shown in Fig.~\ref{Fig_1}, cannot be realized due to the small size of the computational cell having not enough room for domain wall formation.}
\label{Fig_6}
\end{center}
\end{figure}

Note that the presence of plateau on the stress-strain curves in the range $\varepsilon_{xx}^{(1)}<\varepsilon_{xx}<\varepsilon_{xx}^{(2)}$ (see Fig.~\ref{Fig_3}) as well as the linear energy-strain dependence in this range (see Fig.~\ref{Fig_4}) are the features of the two-phase stretching related to the presence of a concave down region on the energy-strain dependence for single translational cell, as described in \cite{SavinDNA}. Let us check if this is the case for the graphene nanoribbon considered here. In Fig.~\ref{Fig_6} we plot the energy-strain dependence for the translational cell containing four atoms (see Fig.~\ref{Fig_2}) subject to periodic boundary conditions. Loading (red solid) and unloading (blue dashed) curves are presented. The concave down regions are clearly seen for both curves. Note that formation of domains with different strain is suppressed by the small volume of the translational cell, that is why the system follows the concave down curve rather than the tangent straight line schematically shown in Fig.~\ref{Fig_1}. Remarkably, the concave down regions are observed in the range of strain where two-phase stretching of the nanoribbons takes place, namely, in the range $\varepsilon_{xx}^{(1)}<\varepsilon_{xx}<\varepsilon_{xx}^{(2)}$ with $\varepsilon_{xx}^{(1)}=0.19$ and $\varepsilon_{xx}^{(2)}=0.375$.

\subsection{Hysteresis Loop}
\label{Sec:Hysteresisloop}

\begin{figure}
\begin{center}
\includegraphics[width=17.8cm]{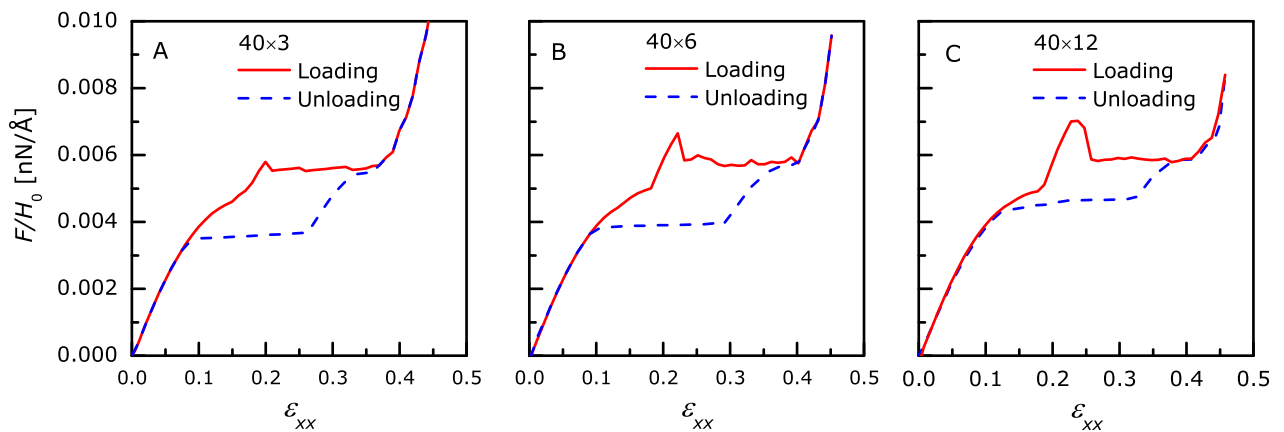}
\caption{Membrane tensile force $F/H_0$ as a function of averaged tensile strain $\varepsilon_{xx}$ during loading (red solid line) and unloading (blue dashed line) of the nanoribbons having width $(A)$ $N=3$, $(B)$ $N=6$, and $(C)$ $N=12$.}
\label{Fig_7}
\end{center}
\end{figure}

It is instructive to study the behavior of the nanoribbon during unloading starting from the strain level $\varepsilon_{xx}>0.375$, when the entire nanoribbon is homogeneously stretched (except for the regions near the clamped nanoribbon ends). The result is shown in Fig.~\ref{Fig_7}, where the dependence of the tensile membrane force as a function of averaged tensile strain is given for loading (red solid line) and unloading (blue dashed line) in the case of nanoribbons with $N=3$, 6, and 12 in $A$, $B$, and $C$, respectively. Overall the nanoribbon shows elastic behavior because at zero tensile force it returns to its original length. However, the unloading curve does not coincide with the loading curve and a hysteresis loop is observed. The area of the loop, which is equal to the energy dissipated due to internal friction, increases with decreasing nanoribbon width.

\begin{figure}
\begin{center}
\includegraphics[width=15cm]{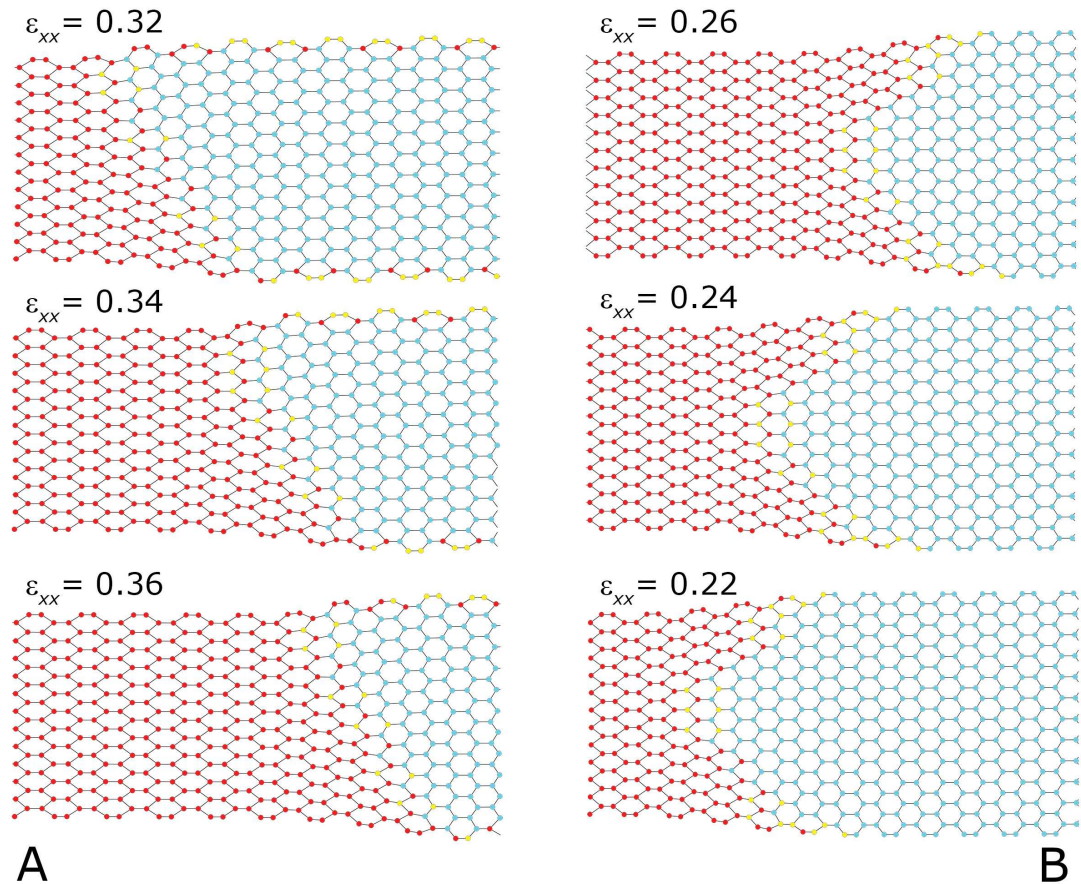}
\caption{DW motion during ($A$) loading and ($B$) unloading of the nanoribbon of width $N=12$. Values of averaged tensile strain are given for each panel. In ($A$) the DW moves to the right, it is tilted, and its propagation begins at the bottom edge of the nanoribbon. In ($B$) the DW moves to the left, it is pointed, and its motion is initiated in the middle of the nanoribbon.}
\label{Fig_8}
\end{center}
\end{figure}

DW dynamics is different for loading and unloading, as can be seen in Fig.~\ref{Fig_8}, and this difference explains why the loading and unloading curves do not coincide. Atoms belonging to the domain with larger (smaller) tensile strain are shown in red (blue). Some atoms are colored in yellow, and they belong to the strongly distorted translational cells located either at the nanoribbon edges or at the DWs. During loading, shown in Fig.~\ref{Fig_8}$A$ (values of the averaged tensile strain are indicated for each panel), the DW moves to the right converting the domain with larger tensile strain into the domain with smaller tensile strain and it is tilted. Motion of this DW begins at the bottom edge of the nanoribbon. During unloading, (see Fig.~\ref{Fig_8}$B$), the DW moves to the left, it has pointed shape, and its propagation starts from the middle of the nanoribbon. The results in Fig.~\ref{Fig_8} are given for the widest studied nanoribbon with $N=12$. A similar picture is observed for the nanoribbons with smaller widths. 

A careful look at Fig.~\ref{Fig_7} reveals that the height of the plateau on the loading curve is almost the same for the nanoribbons with different width and comprises about 0.006~nN/\AA. On the contrary, the plateau level on the unloading branch noticeably increases with increasing $N$, being equal to 0.0035, 0.0039, and 0.0043~nN/\AA$\,$ for $N=3$, 6, and 12, respectively. This result is understandable, taking into account the aforementioned fact that during loading DW motion starts at the nanoribbon edge, while during unloading it starts in the middle of the nanoribbon. Reduction of the unloading plateau level with decreasing $N$ explains the increase in the hysteresis loop area with decreasing $N$.

\begin{figure}
\begin{center}
\includegraphics[width=0.5\linewidth]{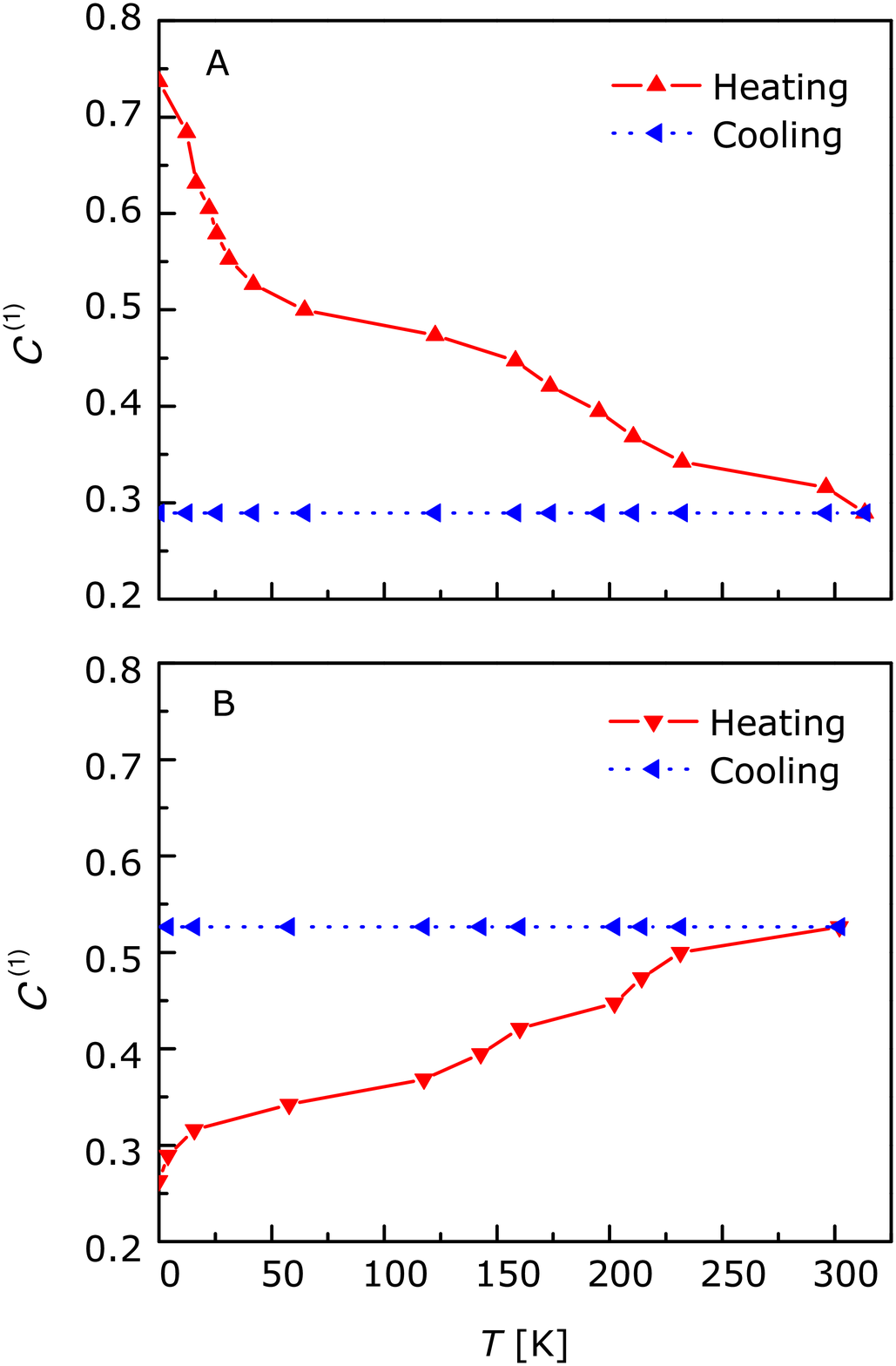}
\caption{Fraction of the domain with smaller local strain, $C^{(1)}$, as a function of temperature for the structure ($A$) on the loading branch ($\varepsilon_{xx}=0.21$, $F/H_0=0.0059$~nN/\AA) and ($B$) on the unloading branch ($\varepsilon_{xx}=0.24$, $F/H_0=0.0035$~nN/\AA). Heating (cooling) is shown by the red solid (blue dashed) line. Results for the narrowest nanoribbon with $N=3$.}
\label{Fig_9}
\end{center}
\end{figure}

\subsection{Temperature effect}
Finally, heating up to 300~K followed by cooling down to 0~K is applied to see the effect of temperature on the DW behavior in the narrowest nanoribbon with $N=3$. Two equilibrium states with DWs are considered: one on the loading branch ($\varepsilon_{xx}=0.21$, $F/H_0=0.0059$~nN/\AA) and the other one on the unloading branch ($\varepsilon_{xx}=0.24$, $F/H_0=0.0035$~nN/\AA), see Fig.~\ref{Fig_7}$A$. Note that both initial structures are on the plateaus of the force-strain curves, which means that they contain domains with local tensile strain equal to $\varepsilon_{xx}^{(1)}=0.19$ and $\varepsilon_{xx}^{(2)}=0.375$. Let $L^{(1)}$ and $L^{(2)}$ being the lengths of the domains with local tensile strain $e_{xx}=\varepsilon_{xx}^{(1)}=0.19$ and $e_{xx}=\varepsilon_{xx}^{(2)}=0.375$, respectively. Note that $L^{(1)}+L^{(2)}=L$, where $L$ is the current length of the nanoribbon. We introduce relative lengths of the domains, $C^{(1)}=L^{(1)}/L$ and $C^{(2)}=L^{(2)}/L$ with $C^{(1)}+C^{(2)}=1$. In Fig.~\ref{Fig_9}, $C^{(1)}$ is shown as a function of temperature for the structure on the loading ($A$) and unloading ($B$) branch. As it can be seen, in Fig.~\ref{Fig_9}$A$, $C^{(1)}$ decreases with temperature on heating and does not change on subsequent cooling. In contrast to that, in Fig.~\ref{Fig_9}$B$, $C^{(1)}$ increases with temperature on heating and similarly to the previous case remains unchanged on cooling. From this result it is clear that not only stretching/unstretching, but also heating can be used to control the DW motion in the two-phase state. We have also demonstrated that the highly stretched nanoribbon survives heating up to room temperature. According to our simulations, fracture of the highly stretched nanoribbon takes place at about 500~K. It is worth noting that the melting temperature of graphene in the absence of loading is 4510~K \cite{Los2015}. 

\section{Discussion and conclusions}

Molecular dynamics study of the uniaxial, strain-controlled tension of graphene nanoribbons with the armchair edges is carried out. It is found that when tensile strain reaches the value of $\varepsilon_{xx}^{(1)}=0.19$, two domains with larger tensile strain of $\varepsilon_{xx}^{(2)}=0.375$ appear near the clamped edges of the nanoribbon. Stretching within the range of the average tensile strain $\varepsilon_{xx}^{(1)}<\varepsilon_{xx}<\varepsilon_{xx}^{(2)}$ occurs through growth of the domain with larger strain in expense of the domain with smaller strain. 

Similar two-phase stretching has been experimentally observed for DNA \cite{Ikai2016133,King20133859,Zhang20133865,Smith1996795}, polypeptides \cite{Afrin20091105}, and some polymer chains \cite{SavinDNA}. This effect is attributed to the presence of a concave down region on the dependence of the potential energy of the unit cell as a function of tensile strain \cite{SavinDNA}, see Fig.~\ref{Fig_6}. However, for the 1D materials, such as DNA and polymer chains, the loading and unloading curves coincide. In the present study of 2D material (graphene nanoribbon) an elastic hysteresis loop is revealed by simulating loading and unloading through the range of strain with two-phase stretching, see schematic Fig.~\ref{Fig_1} for a 1D chain and Fig.~\ref{Fig_7} for the graphene nanoribbon considered here. Structure analysis presented in Fig.~\ref{Fig_8} has shown the difference in the DW profile and kinematics during loading and unloading. Nucleation of DWs and their motion are accompanied by dissipation of the elastic energy. Thus, the nanoribbon can be used as an elastic damper, efficiently converting mechanical strain energy into heat during cyclic loading-unloading through the strain range where domains with larger and smaller strain coexist. 

On the other hand, 2D materials supporting two-phase stretching allow for the new way of creating heterostructures. Fraction of differently strained domains can be controlled by strain-controlled loading or/and by heating, which would result in tuning physical and mechanical properties of the heterostructure.

\ack
The work of I.P.L. was supported by the Russian Foundation for Basic Research, grant no. 17-02-00984-A. E.A.K. thanks the Russian Science Foundation, grant no. 16-12-10175, for their financial support. S.V.D. gratefully acknowledges financial support provided by the Russian Science Foundation, grant no. 14-13-00982.

\section*{References}
\bibliographystyle{iopart-num}
\bibliography{GrapheneDamper}
%\begin{thebibliography}{}

%\end{thebibliography}

\end{document}